# Unraveling the Spin-to-Charge Current Conversion Mechanism and Charge Transfer Dynamics at Interface of Graphene/WS$_2$ Heterostructures at Room Temperature

Rafael O. Cunha[a,*], Yunier Garcia-Basabe[b], Dunieskys G. Larrude[c], Matheus Gamino[d], Erika N. Lima[e,f], Felipe Crasto de Lima[f], Adalberto Fazzio[f], Sergio M. Rezende[g], Antonio Azevedo[g], and Joaquim B. S. Mendes[a]

[a]Departamento de Física, Universidade Federal de Viçosa, 36570-900 Viçosa, Minas Gerais, Brazil
[b]Centro Interdisciplinar de Ciências da Natureza, Universidade Federal da Integração Latino-Americana, 85867-970 Foz do Iguaçu, Paraná, Brazil
[c]Escola de Engenharia, Universidade Presbiteriana Mackenzie, São Paulo 01302-907, Brazil
[d]Departamento de Física, Universidade Federal do Rio Grande do Norte, 59078-900 Natal, Rio Grande do Norte, Brazil
[e]Instituto de Física, Universidade Federal de Mato Grosso, 78060-900 Cuiabá, Mato Grosso, Brazil
[f]Ilum School of Science, Brazilian Center for Research in Energy and Materials (CNPEM), 13083-970 Campinas, São Paulo, Brazil
[g]Departamento de Física, Universidade Federal de Pernambuco, 50670-901 Recife, Pernambuco, Brazil

We report experimental investigations of spin-to-charge current conversion and charge transfer dynamics (CT) at the interface of graphene/WS$_2$ van der Waals heterostructure. Pure spin current was produced by the spin precession in the microwave-driven ferromagnetic resonance of a permalloy film (Py-Ni$_{81}$Fe$_{19}$) and injected into the graphene/WS$_2$ heterostructure through the spin pumping process. The observed spin-to-charge current conversion in the heterostructure is attributed to inverse Rashba-Edelstein effect (IREE) at the graphene/WS$_2$ interface. Interfacial CT dynamics in this heterostructure was investigated based on the framework of core-hole-clock (CHC) approach. The results obtained from spin pumping and CHC studies show that the spin-to-charge current conversion and charge transfer process are more efficient in the graphene/WS$_2$ heterostructure compared to isolated WS$_2$ and graphene films. The results show that the presence of WS$_2$ flakes improves the current conversion efficiency. These experimental results are corroborated by density functional theory (DFT) calculations, which reveal (i) Rashba spin-orbit splitting of graphene orbitals and (ii) electronic coupling between graphene and WS$_2$ orbitals. This study provides valuable insights for optimizing the design and performance of spintronic devices.

# I. Introduction

Scientific research consistently seeks to discover new phenomena and reevaluate existing ones through the lens of advancing technology. In the field of materials physics, spintronics has attracted significant attention from several research groups, paving the way for new discoveries and phenomena, such as spin Hall effect, spin current manipulation and orbitronics.[1–5] The investigation of nanoscale material structures is fundamental to these efforts, particularly in preserving spin memory. Although thin films have traditionally dominated this area, a class of two-dimensional materials, including graphene and transition-metal dichalcogenides (TMD) has emerged as a central point of interest. These materials are very promising, especially in the domain of device fabrication for applications spanning sensors, electronics, spintronics and optoelectronics.[6]

In spintronics, graphene has emerged as an attractive candidate for spin current transport despite its inherent low spin-orbit coupling (SOC), resulting in limited conversion of spin current to charge current.[7–9] However, it has been explored in conjunction with ferromagnetic materials to induce SOC by a proximity effect within the graphene layer.[10] On the other hand, TMDs exhibit strong SOC but moderate spin mobility.[11] In this work, we take advantage of these two characteristics by exploiting a heterostructure composed of graphene and TMD tungsten disulfide ($WS_2$). This combination induces strong spin-orbit coupling in graphene through the proximity effect, facilitating efficient spin current conduction and spin current-to-charge current conversion.[9,12] Exploring the physics of these materials has been greatly facilitated by advances in nanofabrication processes. However, manipulating two-dimensional structures, such as through nanolithography, requires sophisticated techniques.[13] In our research, we employ a simplified fabrication approach for the heterostructure, optimizing time and costs to increase efficiency in large-scale production.

It is known that the proximity effect induced in graphene increases the efficiency of converting spin current into charge current through the inverse Rashba-Edelstein effect (IREE).[12,14] This work aims to elucidate the underlying mechanisms that drive this efficiency increase observed in spin-pumping measurements by microwave-driven ferromagnetic resonance (FMR), at room temperature. Specifically, we investigate the electronic coupling between the S3p states of $WS_2$ and the electronic states of the conduction band of graphene. Our investigation of the charge transfer dynamics in this heterostructure is complemented by ab initio calculations employing density functional theory. The results obtained in this paper provide valuable insights into optimizing the design and performance of spintronic devices based in 2D/2D van der Waals heterostructures.

## II. Sample Preparation and Characterization

The heterostructure comprising a single layer of graphene (SLG) and a layer of tungsten disulfide ($WS_2$), henceforth referred to as $SLG/WS_2$, was synthesized in two steps: first CVD graphene was first transferred onto a $Si/SiO_2$ wafer using a well-established procedure known in the literature,[10,15] then, $WS_2$ was subsequently deposited onto the transferred SLG sheet. This deposition was obtained through the spin coating method at 700 rpm by 60 s, using a solution of $WS_2$ nanoflakes dispersed in ethanol/water (26 mg/L).[16] Figure 1(a) shows an illustrative scheme of the $SLG/WS_2$ heterostructure placed on a $SiO_2$ substrate.

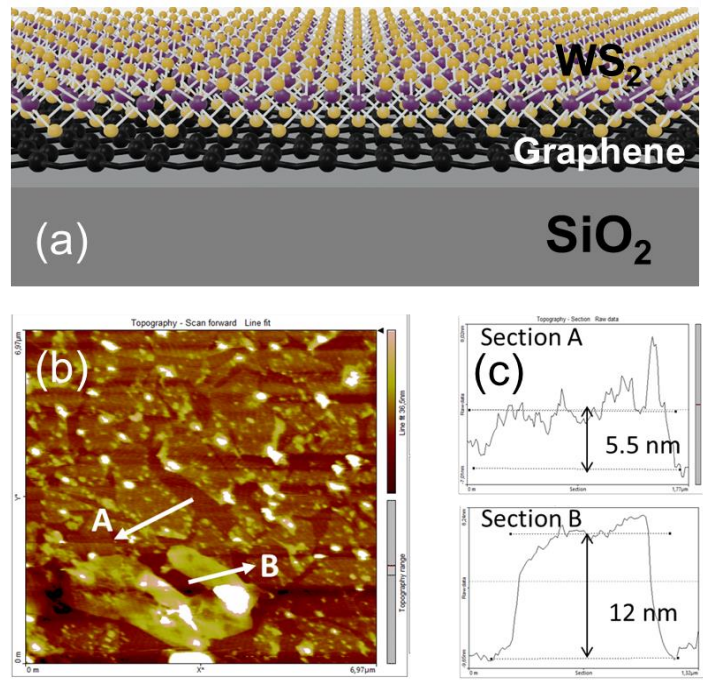

**Figure 1**. (Color online) (a) Illustrative schematic diagram of the $SLG/WS_2$ heterostructure on a $SiO_2$ substrate (grey). Carbon atoms of the SLG are depicted as black dots, while sulfur and tungsten atoms of $WS_2$ are represented by yellow and purple dots, respectively. (b) Atomic force microscopy height image of the $SLG/WS_2$ heterostructure. (c) Height profile in regions A and B of Figure 1(b).

The morphology and thickness of the SLG/WS$_2$ heterostructure were investigated by Atomic Force Microscopy (AFM) microscopy. Figure 1(b) reveals significant spatial heterogeneity in the contact region between the WS$_2$ nanosheets and the graphene layer, with film surface roughness of 4.3 nm. The height profile shows thickness variation ranging from 5.5 nm to 12 nm, indicating the presence of 5-15 layers of WS$_2$ over SLG.

To investigate the crystalline phase of the heterostructure, Raman spectra were collected at room temperature using a WITec Alpha 300R confocal Raman imaging microscope, using a laser line centered at 532 nm with a power of 0.508 mW. Figure 2(a) shows the Raman spectrum SLG/WS$_2$ heterostructure, revealing characteristic vibrational modes typical of graphitic materials, namely the D, G, and 2D peaks.[17] The G-band corresponds to the in-plane optical mode arising from the bond stretching of graphitic sp$^2$ carbon atoms. The D-band originates from the lattice defect and distortion, while the 2D-band is attributed to inter- and intra-valley scattering processes. The two peaks observed at 357 and 423 cm$^{-1}$ (see inset Figure 2(a)) correspond to the in-plane $E_{2g}^1$ and out-plane $A_g^1$ vibrational modes of the 2H-WS$_2$, respectively.[18–20] These values indicate that WS$_2$ has a semiconductor phase.

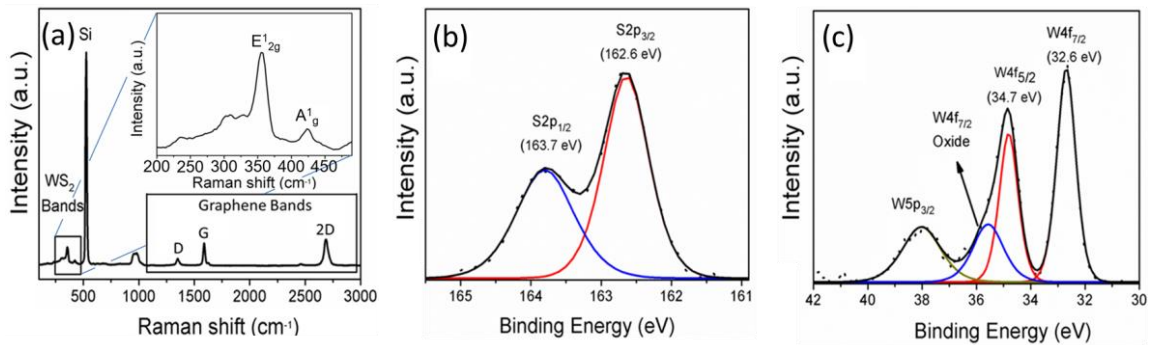

**Figure 2**. (Color online) (a) Raman spectrum of the SLG/WS$_2$ heterostructure. The inset shows $E_{2g}^1$ and $A_g^1$ Raman modes of WS$_2$ in detail. High-resolution XPS spectra of SLG/WS$_2$ heterostructure: (b) S 2p and (c) W 4f core levels. Blue and red lines represent the fittings of XPS spectra.

In addition to structural characterizations Figs. 2(b) and 2(c) show the S2p and W4f high-resolution X-ray Photoelectron Spectra (XPS) spectra of SLG/WS$_2$, respectively. XPS measurements were performed inside an ultra-high vacuum (UHV) chamber with a base pressure of 10$^{-8}$ mbar, using a hemispherical electron energy analyzer (Specs model PHOIBOS 150) with a 45° take-off direction of electrons and pass energy of 25 eV. Photon energy calibration was performed using the Au 4f$_{7/2}$ core level at 84.0 eV of a gold foil. The total energy resolution achieved was 0.76 eV. Fittings of XPS spectra (blue and red lines) were performed using pseudo-Voigt profile functions, with the background correlation performed using a Shirley function. Surface charge effects, manifesting as shifts in electron energy, were monitored using the C (1s) (C-C) photoemission line localized at a binding energy of 285 eV. S 2p XPS spectra (see Figure 2(b)) is characterized by S $2p_{3/2}$ (162.6 eV) and $2p_{1/2}$ (163.7 eV) spin-orbital doublet. On the other hand, the W 4f XPS spectrum (see Figure 2(c)) is characterized by two peaks at 32.6 eV and 34.7 eV, corresponding to the W $4f_{7/2}$ and W $4f_{5/2}$ components, respectively, indicative of the 2H-WS$_2$ phase.[21,22] Additionally, a peak at 35.8 eV corresponding to W $4f_{7/2}$ of W oxidized species is observed in this spectrum.[22]

### III. Spin-pumping Experiments

For the spin-pumping experiments, a 12 nm thick Py (Permalloy, Ni$_{81}$Fe$_{19}$) film was sputter-deposited in an ultra-high vacuum chamber with a base pressure of $4 \times 10^{-8}$ mbar onto both a SLG film and the SLG/WS$_2$ heterostructure, forming islands at the center of each sample, as shown in the schematics of Figs. 3(a) and 3(c), respectively. To ensure ohmic contact, Ti (30 nm)/Au (5 nm) strips were deposited by sputtering on the ends of the samples. Two thin copper electrodes were fixed to the Ti/Au strips by means of silver paint and connected to the nanovoltmeter, as shown in detail in Figs. 3(a) and 3(c). It is important to note that the electrical contacts are not in contact with the Py film to prevent current shunting. Figures 3(b) and 3(d) show the *I-V* curves for the samples SLG/Py and (SLG/WS$_2$)/Py, respectively, demonstrating the ohmic electrical contact between the electrodes and the Ti/Au. The resistance of SLG film is $R = 2.2$ k$\Omega$, significantly higher than the electrical resistance of the SLG/WS$_2$ heterostructure, with $R = 248$ $\Omega$.

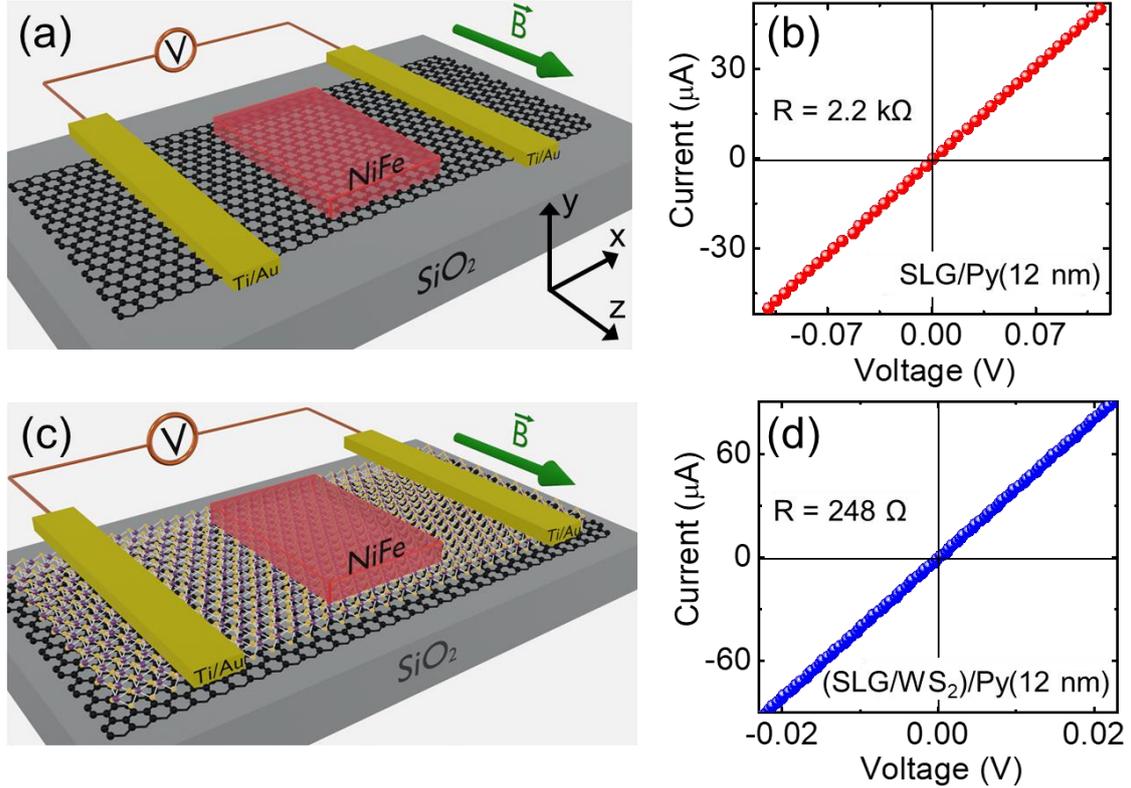

**Figure 3**. (Color online) Illustration of the (a) SLG/Py (12 nm) and (c) (SLG/WS$_2$)/Py (12 nm) structures along with Ti (30 nm)/Au (5 nm) electrodes to ensure ohmic contact. In both samples, the Py film was deposited in the center of the sample without physical contact with the electrodes. The I-V curve demonstrates the ohmic behavior of the contacts between the electrodes and the (b) SLG film and (d) (SLG/WS$_2$) heterostructure.

For the spin pumping experiments, the samples shown in Figs. 3(a) and 3(c) are mounted to the tip of a polyvinyl chloride (PVC) rod and inserted through a hole drilled in the center of the back wall of a rectangular microwave cavity. This cavity operates in the transverse electric ($TE_{102}$) mode at a frequency of 9.4 GHz with a $Q$ factor of 2000. The tip of PVC is placed in a position of maximum rf magnetic field and zero electric field to avoid the generation of galvanomagnetic effects driven by the RF electric field. The microwave cavity is positioned between the poles of an electromagnet so that the sample can be rotated while maintaining the static and rf fields in the sample plane and perpendicular to each other. With this configuration, one can investigate the angular dependence of the FMR absorption spectra. Field scan spectra of the derivative of the microwave absorption $\frac{dP}{dH}$ are obtained by modulating the dc field with the weak ac field, $h_{ac}(t) = h_m \sin(2\pi f t)$, where $f = 1.2$ kHz and $h_m \approx 1.0\ Oe$. This field serves as the reference for lock-in detection of the absorption of the RF field.

Figure 4 shows the FMR absorption spectrum of the Py layer deposited directly on Si substrate, as well as on the SLG film and the SLG/WS$_2$ heterostructure. These measurements were performed with an incident microwave power of 157 mW and a resonance field of $H_R = 1$ kOe. The FMR absorption line has the shape of a Lorentzian derivative with half-width at half-maximum (HWHM) linewidth of $\Delta H_{Py} = 23$ Oe for Py film on de Si substrate, as shown in Figure 4(a). Figure 4(b) shows the absorption line for a sample composed of SLG/Py, without the WS$_2$ sheet. The HWHM linewidth for this sample is $\Delta H_{SLG/Py} = 39$ Oe. The heterostructure compound by (SLG/WS$_2$)/Py has the HWHM linewidth of $\Delta H_{(SLG/WS_2)/Py} = 42$ Oe, as shown in Figure 4(c). These results show that the presence of the single layer graphene and SLG/WS$_2$ heterostructure produces an additional damping due to the spin pumping process[23,24] of $\delta H_{SLG} \equiv (\Delta H_{SLG/Py} - \Delta H_{Py}) = 16$ Oe and $\delta H_{SLG/WS_2} \equiv (\Delta H_{(SLG/WS_2)/Py} - \Delta H_{Py}) = 19$ Oe respectively. Similar effects have been observed in bilayers formed by Py in atomic contact with different materials: normal metals/Py,[25–29] semiconductors/Py,[30–35] graphene/Py,[10,36,37] topological insulators/Py.[38–40]

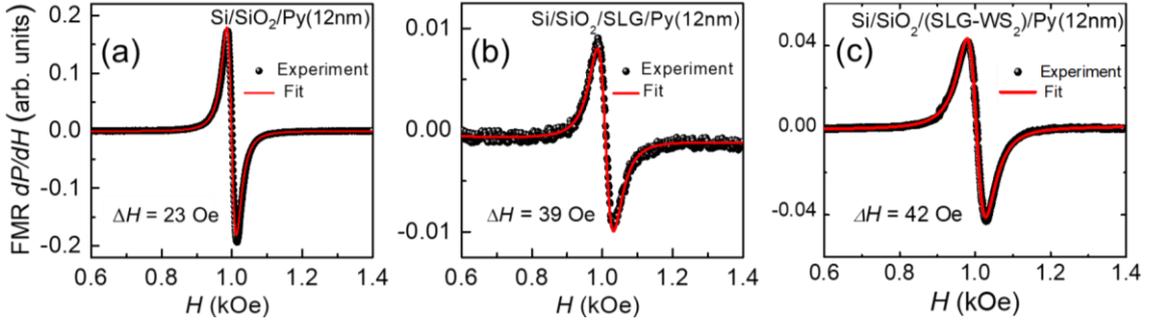

**Figure 4.** (Color online) Panels show the field scan FMR microwave absorption derivative spectra, $\frac{dP}{dH}$ at a frequency 9.4 GHz for three samples: (a) pure 12 nm thick Py film, (b) SLG/Py and (c) (SLG/WS$_2$)/Py, with the magnetic field applied in the film plane.

The precessing magnetization $\vec{M}$ in the Py layer (FM layer) generates a spin current density $\vec{J}_S$, which flows perpendicularly to the interfaces with non-magnetic materials (NM), such as the SLG film or SLG/WS$_2$ heterostructure. This spin current exhibits transverse spin polarization $\hat{\sigma}$, defined by $\vec{J}_S = \left(\frac{\hbar g_{eff}^{\uparrow\downarrow}}{4\pi M_s^2}\right)\left(\vec{M}(t) \times \frac{\partial \vec{M}(t)}{\partial t}\right)$. Here, $g_{eff}^{\uparrow\downarrow}$ is the real part of the effective spin mixing conductance at the interface that takes into account both the spin-pumped and back-flow spin currents,[41] while $M_s$ denotes the saturation magnetization. Expressing the magnetization as $\vec{M}(t) = M_z \hat{z} + (m_x \hat{x} + m_y \hat{y})e^{i\omega t}$, where $m_x, m_y \ll M_z$, the DC spin current density pumped through the FM/NM interface is given in units of angular moment per area per time:

$$J_S(0) = \frac{\hbar \omega g_{eff}^{\uparrow\downarrow} p}{16\pi}\left(\frac{h_{rf}}{\Delta H}\right)^2 L(H - H_r). \tag{1}$$

Here, $\omega$ represents the microwave excitation frequency, $h_{rf}$ is the amplitude of the microwave magnetic field, $L(H - H_r) = \frac{\Delta H^2}{[(H-H_r)^2 + \Delta H^2]}$ is the Lorentzian function, $p$ is the ellipticity factor for the precession cone $p = 4\omega \frac{(H_r + 4\pi M_{eff})}{\gamma}(2H_r + 4\pi M_{eff})^2$, and $\Delta H$ is the FMR linewidth at HWHM. The pure spin current that flows through the FM/NM interface consists of charge carriers with opposite spins moving in opposite directions. As it diffuses into the NM, it creates a spin accumulation given by $J_S(y) = J_S(0)\left\{\frac{\sinh\left[\frac{(t_{NM}-y)}{\lambda_S}\right]}{\sinh\left(\frac{t_{NM}}{\lambda_S}\right)}\right\}$. Here, $t_{NM}$ and $\lambda_S$ respectively denote the thickness and spin diffusion length of the NM layer, such as SLG or SLG/WS$_2$.

In bulk states, the inverse spin Hall effect (ISHE) converts a portion of the spin current density into a transverse charge current density given by $\vec{J}_C = \theta_{SH}\left(\frac{2e}{\hbar}\right)\vec{J}_S \times \hat{\sigma}$, where $\theta_{SH}$ represents the spin Hall angle, quantifying the efficiency of spin-to-charge current conversion, and $\hat{\sigma}$ is the spin polarization. Integrating the charge current density along x and y (as depicted in the inset in Fig 5(a)), yields the spin pumping current, $I_{SP} = \frac{V_{SP}}{R} = w\lambda_S \theta_{SH}\left(\frac{2e}{\hbar}\right)\tanh\left(\frac{t_{NM}}{2\lambda_S}\right)J_S(0)$, where $w$ is the width of the NM material (the distance between electrodes). Similarly, surface states can exhibit analogous phenomena via the inverse Rashba-Edelstein effect (IREE). Through measurements of FMR absorption and spin-pumping voltage, one can extract material parameters by investigating these two distinct phenomena (ISHE and IREE). However, our focus in this work is to explore the influence of WS$_2$ in two-dimensional structures on the enhancement of spin-charge conversion. Thus, we will concentrate on the IREE phenomenon.

To compare the spin pumping response of the SLG/WS$_2$ heterostructure with the responses obtained for SLG, we use the spin pumping current $I_{SP} = \frac{V_{SP}}{R}$, where $R$ is the electric resistance between the two electrodes. $I_{SP}$ is a more significant physical parameter because $V_{SP}$ is proportional to the resistance, which is a parameter dependent on electric contact artifacts. Figure 5(a) shows the current $I_{SP}$ plotted as a function of the applied magnetic field $H$, with a microwave power of 157 mW. As the field sweeps, a *dc* voltage $V_{SP}$ is directly measured at the FMR field value. To achieve this, we employed a nanovoltmeter connected to the Ti/Au electrodes via copper wires. In the case of the SLG/Py sample, the resistance $R$ is measured to be 2.2 k$\Omega$. The SP current is positive for $\phi = 180°$ (blue curve), changes sign with field inversion ($\phi = 0°$, red curve), and vanishes for the field along the sample strip $\phi = 90°$ (black curve). The contributions of galvanomagnetic or spin rectification effects,[26,42] generated by the Py layer itself, can be neglected. This is due to the to the Py film being deposited on the SLG (as well in SLG/WS$_2$ heterostructure) in a manner that isolates it as an island between the electrical contacts, thus preventing any shunting of the charge current. Figure 5(b) shows the field dependence of $I_{SP}$ for the (SLG/WS$_2$)/Py sample, where $R = 248\,\Omega$. Here, one can observe a behavior similar to that of the sample without WS$_2$. Comparing the maximum values of the currents $I_{SP}$ measured at the resonance in Figs. 5(a) and 5(b), both measured at the same rf power $P_{rf} = 157$ mW, we can observe that $I_{SP}^{peak}$ of the (SLG/WS$_2$)/Py sample it is approximately 7 times larger than $I_{SP}^{peak}$ of the SLG/Py sample without the WS$_2$ sheets. This result confirms that the presence of WS$_2$ significantly enhances spin-to-charge interconversion, indicating that the SLG/WS$_2$ heterostructure is substantially more efficient in converting spin current into charge current compared to a single layer of graphene. This large efficiency can be attributed to the large SOC present in transition metals dichalcogenides, such as WS$_2$. A theoretical study on this phenomenon will be discussed later in this work.

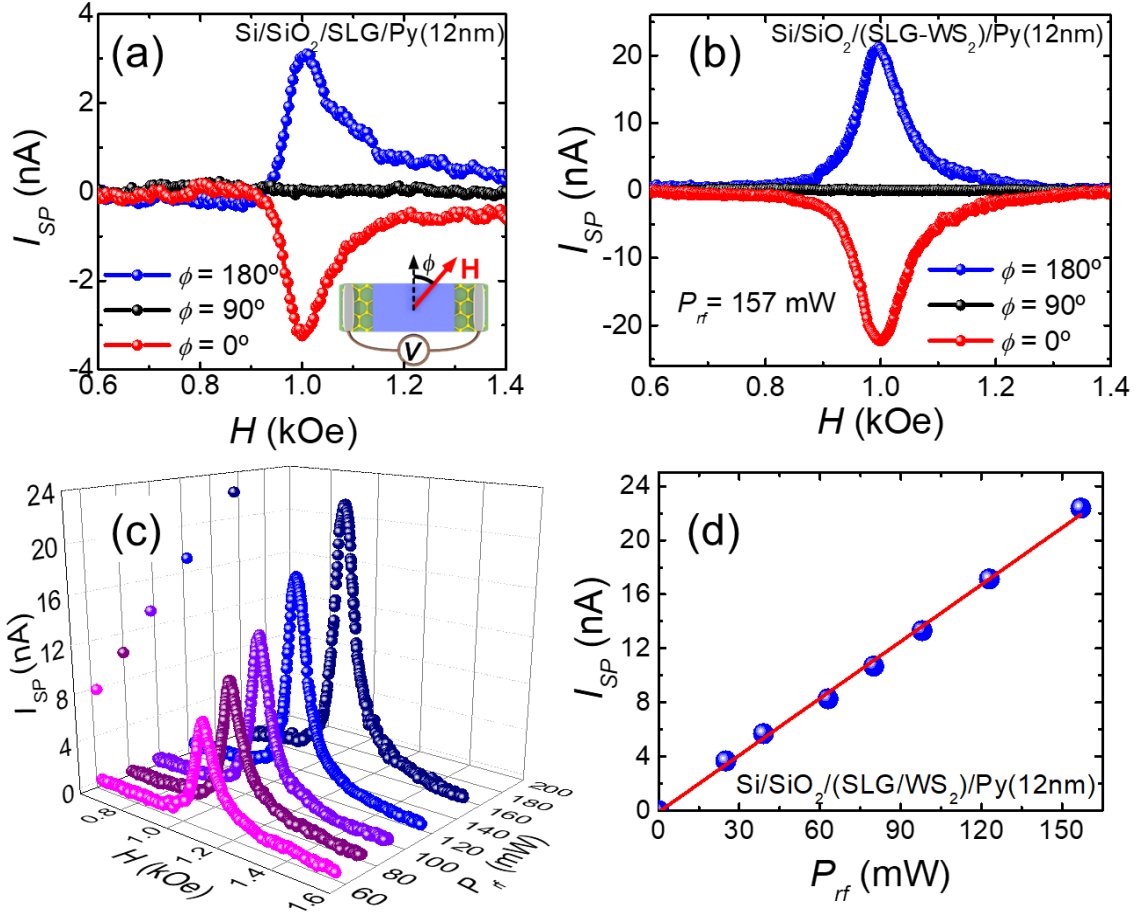

**Figure 5**. (Color online) Field scan of the spin pumping current $I_{SP}$ measured in (a) SLG/Py and (b) (SLG/WS$_2$)/Py at different in-plane field angles, as illustrated in the inset, with an incident microwave power of 157 mW at microwave frequency of 9.4 GHz. (c) Field scans of $|I_{SP}|$ for several values of the incident microwave power for $\phi = 180°$, and (d) peak current value as function of the rf excitation power measured on the (SLG/WS$_2$)/Py sample.

The dependence of the spin pumping current as a function of the applied field, for different rf powers ($P_{rf}$), is shown in Figure 5(c), for $\phi = 0°$. Figure 5(d) shows that the dependence of the maximum value of the spin pumping current ($I_{SP}^{peak}$) on the rf power is in the linear regime of excitation, showing that nonlinear effects are not present in the power range of the experiments.

As previously mentioned, to investigate the enhancement efficiency of the spin-charge conversion process in SLG/WS$_2$ heterostructures, we will focus on the IREE phenomena, particularly suited the two-dimensional characteristics of our samples. We consider that the 3D spin current density $J_S$, generated by the spin pumping process in Py, flows into the SLG film or SLG/WS$_2$ heterostructure, where it undergoes conversion by IREE into a lateral charge current with 2D current density $j_C$. This conversion process is quantified by the IREE coefficient $\lambda_{IREE}$ defined through the relation $j_C = (2\,e/\hbar)\lambda_{IREE}J_S$. Here the charge current density $j_C$ is given by $j_C = I_{SP}^{peak}/w$, where $w = 3.0$ mm is the size of the samples (the distance between electrodes). The spin current density at the interface produced by the FMR spin pumping $J_S$ is calculated using Eq. (1). Combining the current density equations $j_C$ and $J_S$, we derive an expression for the IREE coefficient in terms of the measured charge current peak value:

$$\lambda_{IREE} = \frac{4 I_{SP}^{peak}}{ewf g_{eff}^{\uparrow\downarrow} p_{xz} \left(\frac{h_{rf}}{\Delta H}\right)^2}. \quad (2)$$

The amplitude of the microwave field $h_{rf}$ in Oe is related to the incident power $P_i$, in watt, by $h_{rf} = 1.776(P_i)^{1/2}$, calculated for a microwave cavity made with a shorted standard X-band rectangular waveguide, operating in the $TE_{102}$ mode with $Q$ factor of 2000, at a frequency of 9.4 GHz.[40] For $P_i = 157$ mW this gives $h_{rf} = 0.704$ Oe. $\Delta H$ is the linewidth at HWHM of the sample. The effective spin mixing conductance is given by $g_{eff}^{\uparrow\downarrow} = \left(\frac{4\pi M_S t_{Py}}{\hbar\omega}\right)\delta H$,[10,29,40] which depends on the saturation magnetization $M_S$ and thickness $t_{Py}$ of Py, the increase in the linewidth at HWHM $\delta H$ in Py by the additional damping coming from the spin pumping process due to the contact with the SLG film and SLG/WS$_2$ heterostructure, and the microwave frequency $\omega/2\pi = 9.4$ GHz. The magnetic anisotropy of Py is small, the field of resonance $H_R$ is related to the microwave frequency by the Kittel equation $f = \gamma(H_0 + H_A)^{1/2}(H_0 + H_A + 4\pi M_{eff})^{1/2}$ where $\gamma$ is the gyromagnetic ratio and $4\pi M_{eff}$ is the effective magnetization, which is smaller than the saturation magnetization due to the effect of the perpendicular anisotropy in thin films. Using the measured $H_R = 1.0$ kOe (see Figure 4) and $\gamma = 2.94$ GHz/kOe, corresponding to a g-factor for Py of 2.1, we obtain $4\pi M_{eff} = 9.22$ kG, which is consistent with values for a 12 nm thick Py layer at room temperature.[26,43] The thickness of Py in all samples is $t_{Py} = 12$ nm. The additional damping in SLG/Py and in (SLG/WS$_2$)/Py samples are, respectively, $\delta H_{SLG} = 16$ Oe and $\delta H_{SLG/WS_2} = 19$ Oe, which result in correspondent $g_{eff}^{\uparrow\downarrow}$ values of $2.45 \cdot 10^{18}$ m$^{-2}$ and $2.91 \cdot 10^{18}$ m$^{-2}$.

Finally, using $I_{SP}^{SLG/Py} = (3.11 \pm 0.04)$ nA, $g_{eff}^{\uparrow\downarrow} = 2.45 \cdot 10^{18}$ m$^{-2}$ and $\Delta H_{SLG/Py} = 39$ Oe, we obtain $j_C^{SLG/Py} = 1.04 \times 10^{-6}$ A/m, $J_S^{SLG/Py} = 9.29 \times 10^4$ A/m$^2$ and consequently $\lambda_{IREE}^{SLG/Py} = (0.011 \pm 0.002)$ nm for the SLG/Py sample. Similarly, to the (SLG/WS$_2$)/Py sample, using $I_{SP}^{(SLG/WS2)/Py} = (22,26 \pm 0.07)$ nA, $g_{eff}^{\uparrow\downarrow} = 2.91 \cdot 10^{18}$ m$^{-2}$ and $\Delta H_{(SLG/WS2)/Py} = 42$ Oe, we obtain $j_C^{(SLG/WS2)/Py} = 7.42 \times 10^{-6}$ A/m, $J_S^{(SLG/WS2)/Py} = 9.52 \times 10^4$ A/m$^2$ and $\lambda_{IREE}^{(SLG/WS2)/Py} = (0.078 \pm 0.005)$ nm. The error bar has been incorporated in $\lambda_{IREE}$ by taking into account the variation of $I_{SP}$ measured at $\phi = 0°$ and 180°.

The $\lambda_{IREE}$ value extracted from the sample with WS$_2$ sheets is seven times larger than that observed in the sample without the presence of the TMD, and comparable or even larger than the values reported for several topological insulators.[13,40,44–48] It is important to note that $J_S$ experiences minimal variation with the addition of WS$_2$ over SLG, as this physical quantity essentially depends on the ferromagnetic material (Py), which is the same for both samples. Similarly, the effective spin mixing conductance $g_{eff}^{\uparrow\downarrow}$ displays negligible variation, given that the $\delta H$ linewidth increment is practically identical in both samples. Therefore, the key factor influencing $\lambda_{IREE}$ lies in the magnitude of the charge current $j_C$ resulting from the conversion of $J_S$. Clearly, the presence of WS$_2$, particularly when combined with SLG, significantly increases the spin-to-charge current conversion efficiency of in the investigated heterostructure.

A detailed investigation, as described below, investigated the electronic coupling between the S3p states of WS$_2$ and the electronic states of the conduction band of SLG. This investigation aimed to understand the underlying mechanisms that drive the observed increase in the conversion efficiency of spin current to charge current, as revealed in the spin-pumping measurements.

### IV. Charge transfer dynamics analysis

The electronic coupling between WS$_2$ and graphene was investigated considering the charge transfer (CT) dynamics in SLG/WS$_2$ heterostructures. This transfer was explored through a combination of techniques, including X-ray absorption near edge structure (S K-edge NEXAFS) and Resonant Auger spectroscopy (S–K L$_{2,3}$L$_{2,3}$ RAS), employing the core-hole clock (CHC) approach. Specifically, the interfacial CT dynamics in SLG/WS$_2$ heterostructure are examined for excited electrons from the S 1s to S 3p$_{x,y}$, and S 3p$_z$ electronic states, within the framework of the CHC approach using S–K L$_{2,3}$L$_{2,3}$ RAS spectroscopy.

The CHC approach is element sensitive and dependent on the specific site and orbital, enabling precise investigation of the CT dynamics on the femtosecond timescale. This technique takes advantage of the generation of a core-hole, occurring when a core electron is photoexcited to unoccupied states by X-ray photon absorption. Different decay channels can be observed depending on i) whether the excited electron is still atomically localized during core-hole lifetime or ii) if it is delocalized out of the atom before core-hole refilling. In the first case, two distinct final states can be reached: the spectator two-hole and one electron final states (2h1e), where the excited electron does not participate in the decay process, and the participator one-hole final state (1h), wherein the excited electron actively contributes to the core-hole decay, either through Auger electron emission or core-hole filling. In the second case, the possible final state consists of two holes (2h) in the valence band, reached when the electron is transferred out of the atom during the core-hole lifetime. This process is energetically equivalent to normal Auger decay due to a direct core-level photoionization process and is called the charge transfer (CT-Auger) channel, while the spectator and participator are usually called "Raman" decay channel. The natural lifetime ($\tau_{CH}$) is calculated from the branching ratio of the competing Raman and CT-Auger decay channels, assuming that the two processes are independent and using the $\tau_{CH}$ of the core-excited state as an internal reference. In the CHC approach, depending on the data quality, the accessible CT range is $0.1\tau_{CH} < \tau_{CT} < 10\tau_{CH}$.[49] A schematic representation of CHC approach described above is showed in the supporting information Figure SI-1.

The S K-edge NEXAFS spectrum of SLG/WS$_2$ heterostructure is shown in Figure 6(a). Inside the S K-edge NEXAFS resonance region (below the sulfur ionization threshold of 2474.7 eV), this spectrum is characterized by two main features arising from electron transitions originating from the S1s core level to 3p unoccupied states. The primary feature, appearing at the maximum of the resonance photon energy (2470.8 eV), is attributed to transitions from S1s to $3p_{x,y}$ (in-plane) electronic states, while the second feature, appearing at 2473.1 eV, is attributed to transitions from S1s to $3p_z$ (out-plane) electronic states.[21,50,51]

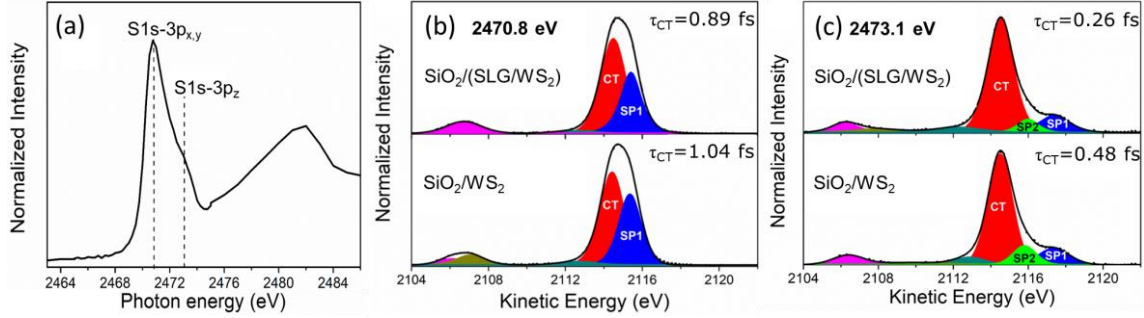

**Figure 6**. (Color online) (a) S K-edge NEXAFS spectra of SLG/WS$_2$ heterostructure collected at 45° incidence angle. The main S1s-3p electronic transitions within the resonance are represented by the dashed lines. S–K L$_{2,3}$L$_{2,3}$ RAS spectra deconvolution in Raman spectator SP1 (blue feature), SP2 (green), and CT (red curve) decay channels collect at photon energies (b) 2470.8 eV corresponding to S1s–3p$_{x,y}$ transition and (c) 2473.1 eV corresponding to S1s–3p$_z$ transition.

Figures 6(b) and 6(c) show S–K L$_{2,3}$L$_{2,3}$ RAS spectra of SLG/WS$_2$ heterostructure obtained at 2470.8 eV and 2473.1 eV photon energies, which correspond to S1s-3p$_{x,y}$ and S1s-3p$_z$ transitions. The S-KL$_{2,3}$L$_{2,3}$ RAS spectra were measured using specific photon energy around the S-K absorption edge with a hemispherical electron energy analyzer (Specs model PHOIBOS 150) with pass energy of 25 eV and 45º take-off direction of Auger electrons. For comparison, the S–K L$_{2,3}$L$_{2,3}$ RAS for a WS$_2$ sample deposited directly on SiO$_2$ substrate, without the single layer of graphene, collected at these same photon energies, were also included in this figure. These decay spectra consist of $^1$S and $^1$D Auger multiplets of the S3p states.[52] The identification of Raman (spectator SP) and CT-Auger decay channels in the RAS spectra is based on their behavior when the incident photon energy is tuned across the core excitation resonance. In Raman decay channels, the kinetic energy of the emitted electron increases when photon energy is tuned across the resonance, while the CT-Auger contribution keeps a constant kinetic energy independent of the energy of the incident photon. The result of this analysis is presented in Figure SI-2.

In order to apply the CHC approach for SLG/WS$_2$ heterostructure, the RAS spectra were deconvoluted in Raman spectator components (SP1 and SP2) and CT-Auger contributions, as shown in Figs. 6(b) and 6(c). The fitting analysis of RAS spectra employed the CASA XPS software package (version 2.3.2), using the Sum Form of Pseudo-Voigt profile functions (linear combination of Gaussian (G) and Lorentzian (L) functions) after subtraction of Shirley-type background. The spectra collected at 2470.8 eV are characterized by CT-Auger contribution appearing at constant kinetic energy of 2114.5 eV (red curve) and spectator contribution SP1 (blue curve) representing the electron localized in $3p_{x,y}$ states. Furthermore, an additional spectator contribution, SP2 (green curve) was observed in the RAS spectra collecting 2473.1 eV corresponding to electrons localized in $3p_z$ states.

The $\tau_{CT}$ values for SLG/WS$_2$ heterostructure and WS$_2$ sample on SiO$_2$ were calculated using $\tau_{CT} = (I_{Raman}/I_{CT\text{-}Auger}) \times \tau_{CH}$, where $I_{Raman}$ and $I_{CT\text{-}Auger}$ are the spectators (SP1 and SP2) and CT-Auger decay integral intensities, respectively. $\tau_{CH}$ represents a core-hole lifetime, which in the case of the S K-edge, is 1.27 fs.[53] Thus, the $\tau_{CT}$ for the SLG/WS$_2$ heterostructure was calculated for S1s–$3p_{x,y}$ (2470.8 eV), and S1s–$3p_z$ (2473.1eV) transitions of 0.89 fs and 0.26 fs, respectively. On the other hand, the $t_{CT}$ values of 1.04 fs and 0.48 fs were obtained for the WS$_2$ sample on SiO$_2$. The electron transfer from these two S3p electronic states is faster in the SLG/WS$_2$ heterostructure compared to the WS$_2$ without SLG system, mainly to electrons excited to $3p_z$ (out-plane) states where is two times faster in SLG/WS$_2$ heterostructure. This result is evidence of electronic coupling between the S3p electronic states of WS$_2$ and the electronic states of the graphene conduction band, favoring the interfacial electron transfer.

These results of charge transfer (CT) analyses performed using the combined S K-edge NEXAFS and S–K $L_{2,3}L_{2,3}$ RAS core level spectroscopy techniques, which indicate that the electronic coupling between the S3p states of WS$_2$ and the conduction band electronic states of SLG favor interfacial electron transfer, corroborate with the results of spin-pumping measurements presented in the previous section. This shows that a heterostructure with the combination of WS$_2$ and SLG makes the system with increased efficiency both in electronic systems and for use in spintronics and combined devices.

### V. Theoretical Interpretation

In order to understand the mechanisms behind the enhancement of the spin-to-charge conversion on SGL/WS$_2$ heterostructure and the electronic coupling between the conduction band states of graphene and the S3p states of WS$_2$, we performed ab initio calculations. Figure 7(a) shows the electronic structure of the SLG/WS$_2$ interface, where the projected orbitals depict the graphene Dirac cone neatly laying within the energy bandgap of WS$_2$. Despite the van der Waals (vdW) nature of the interface, WS$_2$ induces pronounced spin-orbit (SO) effects on the graphene states. This phenomenon induces a spin splitting of the graphene Dirac cone, as depicted in Figure 7(b), similar to a Rashba effect. Such characteristic of the Rashba-like spin-texture is responsible to the experimentally observed spin-charge conversion.[54] In the supporting information we discuss the effect of interface electric field in the increase of the Rashba-spin splitting in the SLG/WS$_2$ interface (Figure SI-3). Here, the effective spin splitting is defined by calculating the mean average over the graphene states:

$$\bar{\Delta}_{SO}(E) = \sum_{n,k} |\langle SLG|n,k\rangle|^2 \, \Delta \, \exp\left(-\frac{(E-\bar{\varepsilon})^2}{\sigma^2}\right)$$

with $\langle SLG|n,k\rangle$ the projected graphene orbitals in the Bloch eigenstates, $\Delta$ is the SOC splitting and $\bar{\varepsilon}$ is the mean energy between the SOC splitted states. Such term accounts not only the SOC splitting, but also the graphene contribution and the density of states. In Figure 7(c) is presented the effective spin splitting. At the Fermi energy we find a value of 1.1 meV while at 1.6 eV above the Fermi energy the splitting increases 10-fold. This can be understood by the superposition of the WS$_2$ states with the graphene pz orbitals. For instance, the projected density of states (Figure 7(d)) shows that at such energy window we have a superposition of W5d states and S3p states with the graphene pz orbitals.

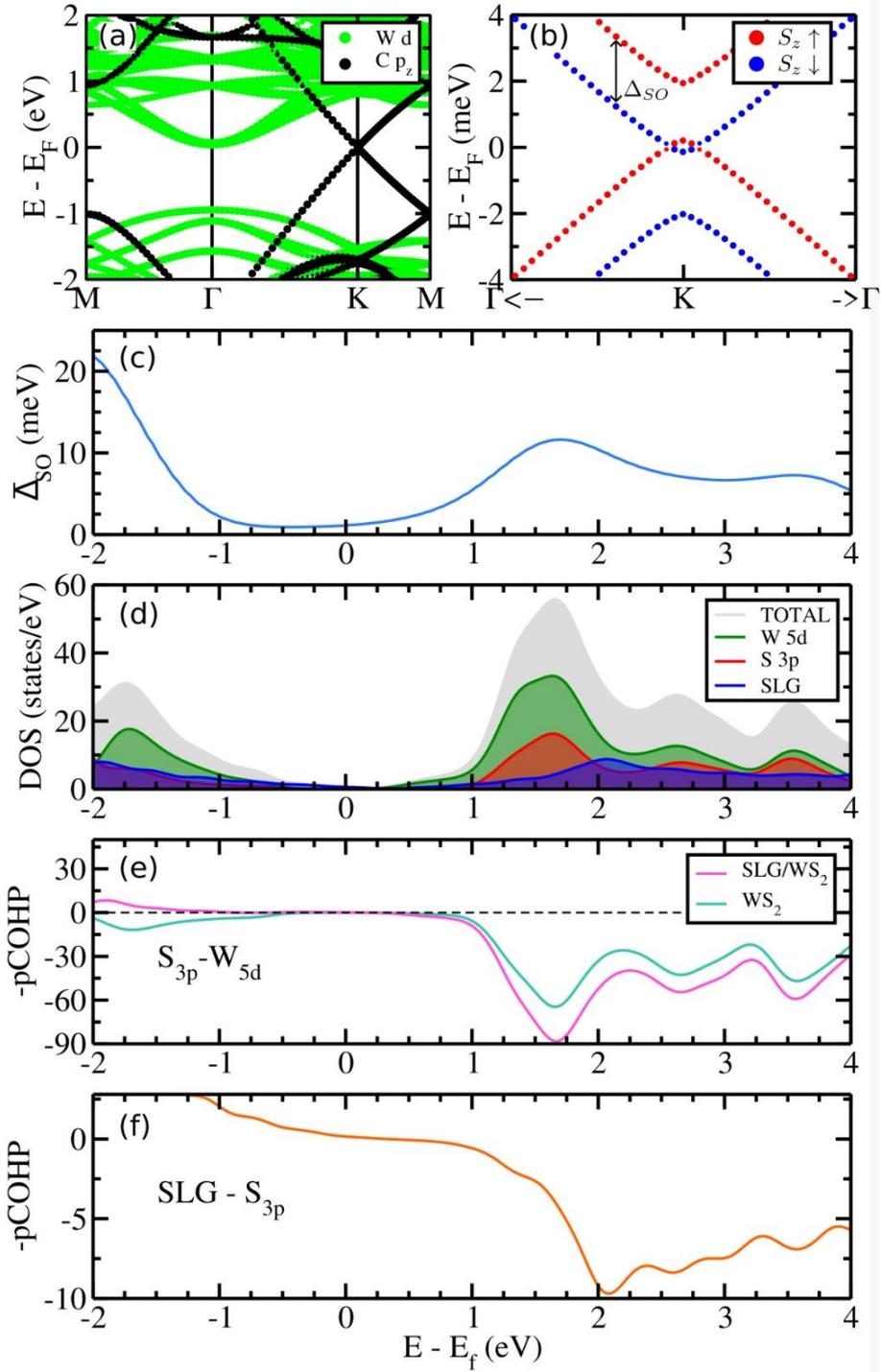

**Figure 7**. (a) Band structure with projected contributions of SLG (black) and WS$_2$ (green) orbitals. (b) Spin projected band dispersion close to the K point. (c) Effective Spin orbit splitting $\bar{\Delta}_{SO}$ of SLG states. (d) Projected density of states. The crystal orbital Hamilton population processed with LOBSTER software (pCOHP) between (e) 3p orbitals of S and 5d orbitals of W, and (f) 3p orbitals of S and SLG states.

Figure 7(d) displays the total and partial density of states (DOS) for SLG/WS$_2$ heterostructure. In the analysis of Figure 7(d), we can observed that W5d states predominantly constitute the conduction band of the SLG/WS$_2$ system. Beyond an energy threshold of 2 eV above the Fermi level, graphene states gradually become more pronounced. Figures 7(e) and 7(f) exhibit the corresponding crystal orbital Hamilton population (-pCOHP) analysis. A significant electronic coupling between S3p states and graphene (Cp) states is observed within the energy range of 2 to 4 eV above the Fermi energy of the system, coinciding with a notable increase in the SOC spin splitting parameter. The significant electronic coupling and the increase of SOC splitting parameter is indicative of the efficient interfacial charge transfer and enhancement of the spin-to-charge conversion process in graphene/WS$_2$ heterostructure.

## V. Conclusion

In this article we report spintronic measurements performed via the ferromagnetic resonance technique on structures composed of SLG/Py and (SLG/WS$_2$)/Py. From these measurements, we extracted the IREE parameters that quantify the spin-to-charge current conversion by the Rashba-Edelstein mechanism, yielding $\lambda_{IREE}^{SLG/Py} = (0.011 \pm 0.002)$ nm and $\lambda_{IREE}^{(SLG/WS2)/Py} = (0.078 \pm 0.005)$ nm, respectively. These results highlight a significant enhancement in the conversion of charge current into spin current by a factor of approximately 7 in the presence of WS$_2$ in contact with the SLG. We attribute this enhancement primarily to the proximity effect observed in graphene. To better understand this enhancement, we carried out a charge transfer study to investigate the electronic coupling between WS$_2$ and graphene. Our results indicate a rapid interfacial electron transfer facilitated by the electronic coupling between the S3p states of WS$_2$ and the conduction band electronic states of SLG within this heterostructure. Additionally, a theoretical binding study was carried out to elucidate the coupling between the electronic states of the conduction band of graphene and the S3p electronic states of WS$_2$, with the findings corroborating the experimental results. Analysis of the density of states shows that W5d states predominantly constitute the conduction band of the SLG/WS$_2$ heterostructure. Furthermore, the spin-orbit effect induced by WS$_2$ in graphene leads to a spin-splitting of the graphene Dirac cone, characteristic of the Rashba effect. In this work we demonstrate that the spin-orbit coupling and the electronic coupling between the states of graphene and WS$_2$ are intrinsically related to each other and the combination of these phenomena is responsible for the enhancement observed in the spin-to-charge current conversion of the SLG/WS$_2$ heterostructure.

**Acknowledgments**

This research is supported by Conselho Nacional de Desenvolvimento Científico e Tecnológico (CNPq), Coordenação de Aperfeiçoamento de Pessoal de Nível Superior (CAPES), Financiadora de Estudos e Projetos (FINEP), Fundação de Amparo à Ciência e Tecnologia do Estado de Pernambuco (FACEPE), Fundação de Amparo à Pesquisa do Estado de São Paulo (FAPESP) (grant 17/02317-2), Universidade Federal de Pernambuco, Multiuser Laboratory Facilities of DF-UFPE, Fundação de Amparo à Pesquisa do Estado de Minas Gerais (FAPEMIG) - Rede de Pesquisa em Materiais 2D and Rede de Nanomagnetismo, INCT of Spintronics and Advanced Magnetic Nanostructures (INCT-SpinNanoMag), CNPq 406836/2022-1. INCT-Materials Informatics, INCT-Nanocarbono, CENAPAD-UNICAMP, and Laboratório Nacional de Computação Científica (ScafMat2).

**Associated Content**

Supporting Information, containing the following sections:

- A- Schematic representation of Core Hole clock approach used to Charge transfer dynamic analysis - Supporting Figure.
- B- The photon energy dependence of electron kinetic energy of Auger decay channels.
- C- Computational Methods.
- D- Effect of interface electric field in the increase of the Rashba-spin splitting.

**Author Information**

*Corresponding Author*

Rafael O. Cunha - Departamento de Física, Universidade Federal de Viçosa, 36570-900 Viçosa, Minas Gerais, Brazil; orcid.org/0000-0002-6039-3892; Email: rafael.cunha@ufv.br

*Authors*

Yunier Garcia-Basabe - Centro Interdisciplinar de Ciências da Natureza, Universidade Federal da Integração Latino-Americana, 85867-970 Foz do Iguaçu, Paraná, Brazil; orcid.org/ 0000-0001-5683-0108; Email: yunier26@yahoo.com.ar

Dunieskys G. Larrude - Escola de Engenharia, Universidade Presbiteriana Mackenzie, São Paulo 01302-907, Brazil; orcid.org/0000-0001-8126-5876; Email: dunieskys.larrude@mackenzie.br

Matheus Gamino - Departamento de Física, Universidade Federal do Rio Grande do Norte, 59078-900 Natal, Rio Grande do Norte, Brazil; orcid.org/0000-0002-0420-7258; Email: mgamino@fisica.ufrn.br

Erika N. Lima - Instituto de Física, Universidade Federal de Mato Grosso, 78060-900 Cuiabá, Mato Grosso, Brazil; Ilum School of Science, Brazilian Center for Research in Energy and Materials (CNPEM), 13083-970 Campinas, São Paulo, Brazil; orcid.org/0000-0002-0670-9737; Email: erika.lima@fisica.ufmt.br

Felipe Crasto de Lima - Ilum School of Science, Brazilian Center for Research in Energy and Materials (CNPEM), 13083-970 Campinas, São Paulo, Brazil; orcid.org/0000-0002-2937-2620; Email: felipe.lima@ilum.cnpem.br

Adalberto Fazzio - Ilum School of Science, Brazilian Center for Research in Energy and Materials (CNPEM), 13083-970 Campinas, São Paulo, Brazil; orcid.org/0000-0001-5384-7676; Email: adalberto.fazzio@ilum.cnpem.br

Sergio M. Rezende - Departamento de Física, Universidade Federal de Pernambuco, 50670-901 Recife, Pernambuco, Brazil; orcid.org/0000-0002-3806-411X; Email: sergio.rezende@ufpe.br

Antonio Azevedo - Departamento de Física, Universidade Federal de Pernambuco, 50670-901 Recife, Pernambuco, Brazil; orcid.org/0000-0001-8572-9877; Email: antonio.azevedo@ufpe.br

Joaquim B. S. Mendes - Departamento de Física, Universidade Federal de Viçosa, 36570-900 Viçosa, Minas Gerais, Brazil; orcid.org/0000-0001-9381-0448; Email: joaquim.mendes@ufv.br

**Data Availability Statement**

The data that support the findings of this study are available from the corresponding authors upon reasonable request.

**Notes**

The authors declare no competing financial interest.